\documentclass[%
 aip,
cp,  
 amsmath,amssymb,
 reprint,%
]{revtex4-2}

\usepackage{graphicx}
\usepackage{dcolumn}
\usepackage{bm}

\usepackage[utf8]{inputenc}
\usepackage[T1]{fontenc}
\usepackage{mathptmx} 

\begin{document}

\title{Mathematical Model of the Stick-Slip Effect for Describing the
	"Drumbeat" Seismic Regime During the Eruption of the Kizimen Volcano in
	Kamchatka}

\author{Parovik Roman} 
\email[Corresponding author: ]{romanparovik@gmail.com}
\affiliation{Vitus Bering Kamchatka State University, Petropavlovsk-Kamchatskiy, Russia 
}
\affiliation{Institute for Cosmophysical Research and Radio Wave Propagation, Far East Branch, Russian Academy of Sciences}

\author{Shakirova Alexandra}%
\affiliation{Vitus Bering Kamchatka State University, Petropavlovsk-Kamchatskiy, Russia 
}
\affiliation{Kamchatka Branch of the Geophysical Survey, Russian Academy of Sciences, Russia 
}
\author{Firstov Pavel}%
\affiliation{Kamchatka Branch of the Geophysical Survey, Russian Academy of Sciences, Russia 
}


\begin{abstract}
During the eruption of the Kizimen volcano in 2010-2013.
There was a uniform squeezing of the viscous lava flow. Simultaneously with its movement, earthquakes with an unusual quasi-periodicity were recorded, the
"drumbeats" mode. In this work, we show that these earthquakes were generated by the movement of the flow front, which was observed for the first time in the practice of volcanological research. We represent the movement of the flow as an intermittent slip with the inclusion of the "stick-slip" mechanism with the initiation of a self-oscillating process. The plausibility of the
phenomenological model at the qualitative level is confirmed by the mathematical model of a fractional nonlinear oscillator.
\end{abstract}

\maketitle

\section{Introduction}

As a rule, the ``drumbeats'' the seismic regime, accompanies the squeezing out
of extrusive domes on andesite and dacite volcanoes of the world \cite{1}-\cite{4}. A
feature of this regime is a well-pronounced quasi-period of the occurrence of
earthquakes from seconds to several tens of minutes with the formation of
multiplets lasting from several hours to tens of days.

To simulate the "drumbeats" mode recorded during the eruption extrusion squeeze in. Soufriere Hills in 1997, considered the simplest model of the process observed in polymer technology \cite{5}. The model assumes that the magma is a Newtonian fluid with a constant viscosity independent of the shear rate. When the melt is squeezed out of the reservoir at a constant speed into a channel with a small diameter, then the movement takes on an intermittent character only when the material flow rate is within a certain range. Physical modeling showed that the oscillations are caused by the boundary conditions of the contact of the melt with the wall. For oscillations to occur, a process is needed when energy can be alternately accumulated and released. In this case, the magma will alternately stick and slide along the channel wall, which corresponds to the "stick-slip" mechanism. The simulation results showed that an increase in the oscillation period is due to a change in the volume or length of the channel, as well as an increase in the viscosity of magma. Whereas, an increase in the bulk modulus or channel radius led to a decrease in the cycle period.

During the eruption of the Tungurahua volcano (Andes, Ecuador) in 2015, the
``drumbeats'' regime was recorded before the extrusion dome appeared \cite{6}. In this case, two mechanisms have been proposed for the occurrence of the drumbeats mode: (1) the process of degassing followed by migration to the surface of the flow of gases; (2) the process of destruction of the geomedium due to the movement of magma. However, the preference is given to the mechanism of the ``drumbeats'' regime due to the movement of the fluid/gas flow along the network of cracks that appear ahead of the gradually rising magma front. However, this model does not provide an explanation for the quasiregularity of the process.

From the above, we can conclude that the occurrence of multiplets of earthquakes in the ``drumbeats'' mode is a unique natural process, the mechanism of which has not yet been fully understood. Understanding the generation process of quasiregular earthquakes requires a search for suitable mathematical models. In this paper, we consider the relationship of the seismic regime ``drumbeats'' with the dynamics of lava flow and propose a mathematical model of its occurrence.

\section{STICK-SLIP MATHEMATICAL MODEL}

At the front of the lava flow, separate lava blocks $m$ are formed (we take them in the form of a parallelepiped), which move under the action of an active core moving at a constant velocity $v$. In this case, the block $m$ can be considered as the right element of the system, moving due to the forces of noncumulative friction ${F_{nec}}$ with a falling characteristic depending on the speed (Fig. 1). Naturally, elements with a stiffness coefficient $k$  and a viscous resistance coefficient $b$ should be included between the active core and individual blocks.
\begin{figure}[h!]
\centering	
\includegraphics[scale=1]{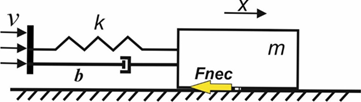}
\caption{Mechanical model of the generation of \guillemotleft{}drumbeats'' seismic regime, which occurs when a lava flow}
\end{figure}

The mechanical model presented in Figure 1 describes the stick-slip effect. The stick-slip effect arises in problems of the movement of a load on a spring on a solid surface, taking into account friction and adhesion, which leads to adhesion of the load to the surface at certain times \cite{7}. When the load moves along the surface, jumps of the load are observed, which are due to adhesion and separation from the surface, quasi-periodic sliding occurs. The magnitude of these jumps depends on the strength of adhesion and the stiffness of the spring.

Also stick-slip effect may be laid in the foundation of the mechanical model of the earthquake in the subduction zone of lithospheric plates \cite{8}. Consider a mathematical model of the stick-slip effect.

Consider the following Cauchy problem \cite{9}:

\begin{equation}
\frac{{{d^2}x\left( t \right)}}{{d{t^2}}} + b\frac{{dx\left( t \right)}}{{dt}} +
{\omega ^2}x\left( t \right) = \upsilon t + c\sum\limits_{n = 1}^7 {{a_n}\sin
	\left( {nx\left( t \right)} \right)} ,x\left( 0 \right) = {x_0},\dot x\left( 0
\right) = {y_0},
\label{eq1}
\end{equation}
where $x\left( t \right)$ is the displacement function,  $t \in \left[ {0,T}
\right]$ -- time $T$ is the simulation time, $\upsilon  = v{\omega ^2}$, $v$ is the displacement speed $\omega $ is the natural frequency, $b$ -- coefficient of friction, $c = \frac{{{U_p}}}{{{x_p}m{\omega ^2}}}$  -- surface adhesion energy, $m = {k \mathord{\left/{\vphantom {k {{\omega ^2}}}} \right. \kern-\nulldelimiterspace} {{\omega ^2}}}$- effective mass of the oscillator, $k$ -- stiffness coefficient, ${U_p}$  and ${x_p}$  -- the depth and width of the potential well, ${a_n} = 2n\int\limits_0^1 {\frac{{\cos \left( {\pi n\tau }\right)d\tau }}{{{{\cosh }^2}\left( {\pi \tau } \right)}}} $ are the coefficients of the expansion of the Fourier series, ${x_0}$  and ${y_0}$ are given constants that determine the initial condition of the oscillatory system (1).

The mathematical model of the stick-slip effect is a nonlinear oscillator with
in the approximation of equally alternating potential wells. Falling into a
potential well means that the displacement rate is close to zero.

Let's show on a qualitative level that this mathematical model can be applied to the description of the seismic regime "drumbeats". The solution to problem (1) was obtained by the numerical Runge-Kutta method of the 4th order in the environment of symbolic computer mathematics Maple 2021.

\section{DATA AND ANALYSIS}

The analysis of the drumbeats seismic regime was carried out on the basis of seismic data from KZV station (Fig. 2, inset) of the Kamchatka Branch of the
Geophysical Survey of the Russian Academy of Sciences (KBGS RAS) \cite{10}, \cite{11} on the vertical SHZ component. KZV was installed in 2009, 2.5 km from the top of the Kizimen volcano, and in the period 2011-2012 it worked almost without failures and malfunctions, which is of particular importance for us to study the quasi-periodicity of the earthquake occurrence process. The work with seismic records was carried out in the DIMAS seismic record processing program \cite{12}.
\begin{figure}[h!]
\centering
\includegraphics[scale=1]{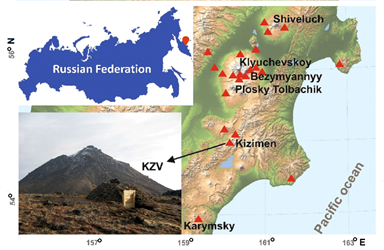}
\caption{Location of the Kizimen volcano and seismic stations in the central part of the Kamchatka Peninsula. The upper inset shows a map of the Russian Federation; the lower inset shows a general view of the Kizimen volcano and the bunker where the KZV seismic station is located}
\end{figure}

Let us consider the change in the nature of the generation of earthquakes in the "drumbeats" mode in the periods of March and October 2011, for which mathematical modeling was carried out. In March (Fig. 3c), a lava flow began to form with a maximum thickness of $h$ = 70 m and a length of 1 km \cite{13}.

\begin{figure}[h!]
\centering
\includegraphics[scale=1.2]{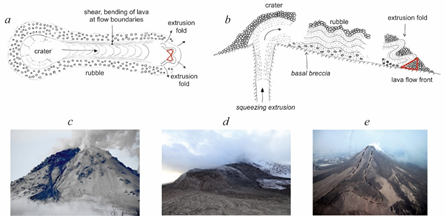}
\caption{Cross-section along the length of the Rocche Rosse lava flow (Lipari Island, Italy) with generalized layering schemes (a) and a schematic plan, section and details of the main structures of block lava on the island. San Pietro (Sardinia, Italy) (b) \cite{14}. The red lines indicate the presumptive division of the lava flow front into separate blocks. Lava flow view on March 4, 2011 (c) and September 14, 2011 (d) and March 15, 2012 (e).}
\end{figure}

During this period, KZV recorded a multiplet of earthquakes lasting 12 days
(March 11-22). On March 20, the average period of earthquake registration was
$T$= 14 s (the average frequency of earthquakes was $\bar f$ = 3.7 min$^{-1}$). The average amplitude of earthquakes in the selected fragment of the seismic record in Fig. 3a was equal to $\bar A$ = 4.5 $\mu{}$m/s (Fig. 4a).
\begin{figure}[h!]
\centering
\includegraphics[scale=1]{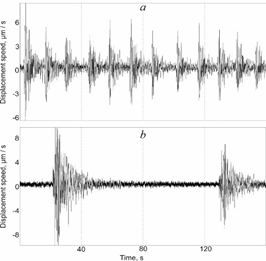}
\caption{Seismic mode ``drumbeats'', registered on 03/20/2011 at KZV s/st, SHZ channel with time countdown 08:23:08 (a) and 10/05/2011 at KZV s/st, SHZ channel with 14:48 time countdown: 37 (b)}
\end{figure}

After analyzing and comparing the structures of the seismic mode multiplets
"drumbeats" in March and October 2011 and their dynamic parameters with the
dynamics of the lava flow, it was concluded that the more powerful tongue of the lava flow generated earthquakes with a lower frequency of registration, but a higher amplitude. Conversely, a thin lava flow, which was characterized by a high velocity of movement, generated more frequent earthquakes with a lower amplitude. The results obtained give grounds to propose the following mathematical model (1) for the generation of the "drumbeats" seismic regime.

Consider the results of modeling using the parameters of the lava flow in March 2011 (Fig.4). The values of the parameters of the model (1) are chosen as follows:
\[
b = 0.6, = {\rm{233}}{\rm{.14}}{\rm{,}}\upsilon  = 16.72,\omega  = 409,m =
{10^5},v = {10^{ - 4}},{x_0} = {y_0} = 0.
\]

For numerical analysis, the number of nodes of the computational grid coincides with the amount of experimental data $N = 18235,$the decoding step is $\tau  = {1 \mathord{\left/{\vphantom {1 {128}}} \right. \kern-\nulldelimiterspace} {128}}$.
\begin{figure}[h!]
\centering
\includegraphics[scale=1]{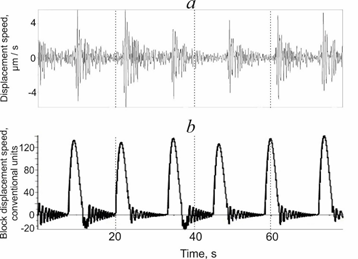}
\caption{Seismic mode ``drumbeats'' recorded on 03/20/2011 with time reference 08:23:53 (a) and block displacement velocity function curve (b) based on lava flow parameters (1)}
\end{figure}

Consider the results of modeling using the parameters of the lava flow in
October 2011 (Fig. 5). The values of the parameters of the model (1) are chosen as follows:
\[
b = 0.1, = 572.{\rm{48}}{\rm{,}}\upsilon  = 5.93,\omega  = 243.7,m = {10^5},v = {10^{ - 4}},{x_0} = {y_0} = 0,N = 112504,\tau  = {1 \mathord{\left/{\vphantom {1 {128}}} \right.\kern-\nulldelimiterspace} {128}}.
\]
\begin{figure}[h!]
\centering
\includegraphics[scale=1]{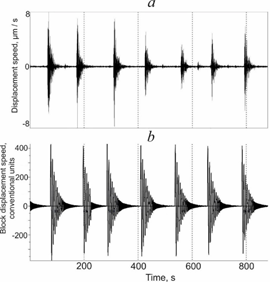}
\caption{Seismic mode ``drumbeats'' recorded on 5/10/2011 with time reference 15:42:17 (a) and block displacement velocity function curve (b) based on lava flow parameters}
\end{figure}

Interpreting the results obtained, we can say that this model in the first
approximation describes the process of generation of quasiperiodic earthquakes of the ``drumbeats'' mode. As can be seen from Fig. 4 and 5, the freezing coefficient of the block $c$, friction $b$ and the speed of movement of the block $v$  have the greatest influence on the oscillations. With an increase in $c$, an increase in the energy of earthquakes occurs, since it is necessary to apply a large force in order to overcome the "freezing" of the block at the front of the lava flow.
	
\section{Conclusion}

The results of mathematical modeling showed that with an increase in the
thickness of the front of the Kizimen volcano lava flow, the friction at the
contact between the base of the flow and the underlying rock increased. The
pressure of the squeezing lava led to the ``breakdown'' of individual blocks of the lava flow front, their slipping and stopping, thus the model of generation of quasiperiodic earthquakes is consistent with the ``stick-slip'' model. More drumbeats in multiplets with lower A were the result of a decrease in sticking, while fewer earthquakes with higher A were characterized by an increase in sticking, which increased with increasing lava flow front strength.

Estimated calculations performed for two examples showed that the model of a
fractional nonlinear oscillator provides self-oscillatory motion of the lava flow front block with a frequency close to the observed one. This is confirmed by the coincidence of the results of mathematical modeling with a fractional nonlinear oscillator on a qualitative level with experimental data.

Based on the simulation results using the known physical parameters of the lava flow, it can be said that the freezing coefficient of the block c had the
greatest influence on the oscillations. To overcome the "freezing" of the block at the front of the lava flow, it was necessary to apply force. The more freezing we observed, the more force needed to be applied. Consequently, with an increase, there was an increase in the energy of earthquakes.

As a continuation of the work, it makes sense to consider a generalization of
the mathematical model (1) to the case of taking into account the memory effect \cite{15} and; however, in this case, the computational complexity of the problem that must be solved by supercomputers will sharply increase.

It is also of practical interest to study the amplitude-frequency and
phase-frequency characteristics of the stick-slip effect by analogy with \cite{16}.

\section{FUNDING}

The work was supported by Ministry of Science and Higher Education of the
Russian Federation, research topic \guillemotleft{}Natural disasters of Kamchatka -- earthquakes and volcanic eruptions\guillemotright{} No.
AAAA-A19-119072290002-9. The data used in the work were obtained with large-scale research facilities \guillemotleft{}Seismic infrasound array for monitoring Arctic cryolitozone and continuous seismic monitoring of the Russian Federation, neighbouring territories and the world\guillemotright{}
(https://ckp-rf.ru/usu/507436/,http://www.gsras.ru/unu/)

\begin{acknowledgments}
R. Parovik thanks prof. S. Rekhviashvili for valuable comments on the
formulation of the stick-slip model. A. Shakirova. and P. Firstov would like to thank V. Garbuzova for prompting them to manifest the seismic regime "drumbeats" accompanying the eruption of the Kizimen volcano.
\end{acknowledgments}


\begin{thebibliography}{9}
\bibitem{1}	D. Sherrod and W. Scott and P. Stauffer, Volcano rekindled: The renewed eruption of Mount St. Helens, 2004-2006 (U.S. Geological Survey Professional Paper 1750, 2008) 856 p.
\bibitem{2}	J. Power and J. Coombs and M. Freymueller, The 2006 eruption of Augustine volcano, Alaska (U.S. Geological Survey Professional Paper 1769-1, 2010) 667 p. 
\bibitem{3}	J. E. Kendrick and Y. Lavalle and T Hirose and G. Di Toro and A. J. Hornby and S. De Angelis end D. B. Dingwell, Volcanic drumbeat seismicity caused by stick-slip motion and magmatic frictional melting. Nature Geoscience 7, 438-442 (2014). 
\bibitem{4}	A. Shakirova and P. Firstov, Observation of the seismic mode «drumbeats» on volcanoes of the world and Kizimen volcano (Russia). E3S Web of Conferences 127, 03004 (2019).  
\bibitem{5}	R. Denlinger and R. Hoblitt, Cyclic eruptive behavior of silicic volcanoes. Geology 27(5), 459–462 (1999)
\bibitem{6}	A Bell and S. Hernandez and H. Gaunt and P. A. Mothes and M. Ruiz and D. Sierra and S. Aguaiza, The rise and fall of periodic 'drumbeat' seismicity at Tungurahua volcano, Ecuador. Earth and Planetary Science Letters, 475, 58–70 (2017).
\bibitem{7}	E. G. Daub and J. M. Carlson, Stick-slip instabilities and shear strain localization in amorphous materials. Physical Review E, 80(6), 066113 (2009).
\bibitem{8}	Ch. H. Scholz, The mechanics of earthquakes and faulting (Cambridge university press, 2002) 471 p.
\bibitem{9}	S. Rekhviashvili, Dimensional phenomena in condensed matter physics and nanotechnology (Nalchik: KBNTs RAN, 2014) 250 p. 
\bibitem{10}	V. N. Chebrov and D. V. Droznin and Yu. A. Kugaenko, and V. I. Levina and S. L. Senyukov and V. A. Sergeev and Yu. V. Shevchenko and V. V. Yashchuk, The system of detailed seismological observations in Kamchatka in 2011. J. Volcan. Seismol 7, 16-36 (2013).
\bibitem{11}	A. Yu. Chebrova and A. S. Chemarev and E. A. Matveenko and D. V. Chebrov, Seismological data information system in Kamchatka branch of GS RAS: organization principles, main elements and key function. Geophysical Research 21, 66-91 (2020).
\bibitem{12}	D. Droznin and S. Y. Droznina, Interactive DIMAS program for processing seismic signals. Seism. Instr. 47(3), 215–224 (2011).
\bibitem{13}	V. Dvigalo and L. Mekekestsev and A. Shevchenko and I. Svirid, The 2010–2012 eruption of Kizimen Volcano: The greatest output (from the data of remote-sensing observations) for eruptions in Kamchatka in the early 21st century part I. The November 11, 2010 to December 11, 2011 phase.   Journal of Volcanology and Seismology 7, 345–361 (2013).
\bibitem{14}	A. Harris and S. Rowl and N. Villeneuve and T. Thordarson, Pāhoehoe, ‘a‘ā, and block lava: an illustrated history of the nomenclature. Bull Volcanol 17, 2–34 (2017).
\bibitem{15}	R. Parovik, On a credit oscillatory system with the inclusion of stick-slip. E3S Web of Conferences 7, 1-5 (2016).
\bibitem{16}	R. I. Parovik, Amplitude-frequency and phase-frequency performances of forced oscillations of a nonlinear fractional oscillator. Technical Physics Letters 45(7), 660-663 (2019).


\end{thebibliography}

\end{document}